\documentclass[grl]{agu2001}
\usepackage[dvips]{graphicx}
\lefthead{HELMSTETTER and SORNETTE}
\righthead{B{\aa}th's law, Gutenberg-Richter law and 
Aftershock Properties}	
\authoraddr{Agn\`es Helmstetter,  Institute of
Geophysics and Planetary Physics, University of California, Los Angeles,
California.  (e-mail: helmstet@moho.ess.ucla.edu)}
\authoraddr{Didier Sornette, Department of Earth and Space Sciences and Institute of
Geophysics and Planetary Physics, University of California, Los Angeles,
California and Laboratoire de Physique de la Mati\`{e}re Condens\'{e}e,
CNRS UMR 6622 and Universit\'{e} de Nice-Sophia Antipolis, Parc Valrose,
06108 Nice, France (e-mail: sornette@moho.ess.ucla.edu)}

\input{update.tex}
\begin{document}
\title{B{\aa}th's law Derived from the Gutenberg-Richter law and from
Aftershock Properties}
\author{Agn\`es Helmstetter$^1$ and Didier Sornette$^{1,2,3}$}
\affil{$^1$ Institute of Geophysics and Planetary Physics, 
University of California, Los Angeles,        California 90095-1567\\
$^2$ Department of Earth and Space Sciences, University of California,
Los Angeles,       California 90095-1567 \\
$^3$    Laboratoire de Physique de la Mati\`{e}re Condens\'{e}e, CNRS UMR 6622
Universit\'{e} de Nice-Sophia Antipolis, Parc Valrose, 06108 Nice, France}
\newcommand{\be}{\begin{equation}}
\newcommand{\ee}{\end{equation}}
\newcommand{\ba}{\begin{eqnarray}}
\newcommand{\ea}{\end{eqnarray}}
\newenvironment{technical}{\begin{quotation}\small}{\end{quotation}}

\begin{abstract}
The empirical B{\aa}th's law 
states that the average difference in magnitude between a mainshock and
its largest aftershock is $1.2$, regardless of the mainshock magnitude. 
Following {\it Vere-Jones}' [1969] and {\it Console et
  al.} [2003], we show that the origin of B{\aa}th's law is to be
found in the selection procedure used to define mainshocks and aftershocks
rather than in any difference in the mechanisms controlling the
magnitude of the mainshock and of the aftershocks. 
We use the ETAS model of seismicity, which provides a more realistic 
model of aftershocks, based on (i) a universal Gutenberg-Richter (GR) 
law for all earthquakes, and on (ii) the increase of the number of aftershocks 
with the mainshock magnitude. Using numerical simulations of the ETAS model, 
we show that this model is in good agreement with B{\aa}th's law in a certain range of
the model parameters.
\end{abstract}

\begin{article}

\section{Introduction}

B{\aa}th' s law [{\it B{\aa}th}, 1965] predicts that the average 
magnitude difference $\Delta m$  between
a mainshock and its largest aftershock is $1.2$,
independently of the mainshock magnitude. 
Many studies have validated B{\aa}th's law, with however large
fluctuations of  $\Delta m$ between 0 and 3 from one
sequence to another one [e.g. {\it Felzer et al.}, 2002; 
{\it Console et al.}, 2003]. In addition to providing
useful information for understanding earthquake processes,
B{\aa}th's law is also important from a societal view point as it gives
a prediction of the expected size of the potentially most
destructive aftershock that follows a mainshock.

\section{Vere-Jones' interpretation of B{\aa}th's law}

B{\aa}th's law is often interpreted as an evidence that mainshocks
are physically different from other earthquakes and have a different
magnitude distribution [e.g. {\it Utsu}, 1969].
In contrast, {\it Vere-Jones} [1969] offered a 
statistical interpretation, elegant in its simplicity, which consisted 
in viewing the magnitudes of the mainshock and largest aftershock
as the first and second largest values of a set of independent
identically distributed (iid) random variables distributed according
to the same GR distribution $P(m)\sim 10^{-bm}$.
If the same minimum threshold $m_0$ applies for both aftershocks and
mainshocks, this model predicts that $\Delta m$ has the same density 
distribution $P_{\Delta m} (\Delta m) \sim 10^{-b  \Delta m}$
as the GR distribution of the sample
[{\it Vere-Jones, 1969}] with a mean $\langle \Delta m \rangle$ equal to
$1/(b \ln 10) \approx 0.43$ for $b \approx 1$. Thus, rather than
a distribution peaked at $\Delta m \approx 1.2$, {\it  Vere-Jones}'
interpretation predicts an exponential distribution with an average
significantly smaller than B{\aa}th's law value $\approx 1.2$.
Such discrepancies have been ascribed to different magnitude
thresholds chosen for the definition of mainshocks and largest
aftershocks and to finite catalog size effects
[{\it Vere-Jones}, 1969; {\it Console et al.}, 2003].
Improved implementation of {\it Vere-Jones}' model by
{\it Console} et al. [2003], taking into account the fact
that the minimum threshold for aftershock magnitudes is smaller
than for mainshocks, has shown that this model provides
a much better fit to the data,  but that there is still a minor
discrepancy between this model and the observations.
The results of [{\it Console et al.}, 2003]
for a worldwide catalog and for a catalog of seismicity of New Zealand
are not completely explained by this model, the observed value of 
$\langle \Delta m \rangle$ being still a little larger than predicted.
{\it Console et al.} [2003] interpret this result as possibly due to
``a change in the physical environment before and after
large earthquakes'' but they do not rule out the existence
of a possible bias that may explain the discrepancy between
their model and the observations.
We propose in section \ref{ETAS} a simple statistical
interpretation of B{\aa}th's law, which can explain this discrepancy 
without invoking any difference in the mechanisms controlling the
magnitude of the mainshock and of the aftershocks.

Notwithstanding the appealing simplicity of {\it Vere-Jones}'
interpretation and its success to fit the data, this model does not 
provide a realistic model of aftershocks, and misses some important
properties of seismicity. In particular, it does not take into account
the fact that aftershocks represent only a subset of the whole seismicity,
which are selected as events that occurred within a space-time window 
around and after a larger event, called the mainshock, which is supposed to
have triggered these earthquakes.
We first consider as a mainshock only the largest earthquake of a catalog 
of $N$ events which have independent magnitudes drawn according to the GR 
law $10^{-b (m-m_0)}$ with a minimum magnitude $m_0$.  
Only a small subset of size $N_{\rm aft}$  of the whole catalog
occur in the specified space-time window used for aftershock selection. 
The largest event in the whole catalog has an average magnitude given by 
$\langle m_{M} \rangle \approx m_0 + (1/b) \log_{10} N$. 
Let us sort the magnitudes of all events in the catalog by descending order: 
$m_1 > m_2 > ... > m_N$.
The largest aftershock, within the subset of aftershocks of size 
$N_{\rm aft}$, has an  expected overall rank equal to $\approx N/N_{\rm aft}$ .
Using the distribution of the magnitude difference $m_1-m_j$ between 
the largest earthquake (with rank equal to 1) and the event of rank $j$ 
given by [{\it Vere-Jones}, 1969] and assuming $ N \gg  N_{\rm aft} \gg 1$,
the average magnitude difference between the mainshock and its largest
aftershock is thus given by 
 \be
\langle \Delta m \rangle =\langle m_M-m_A \rangle 
\approx {1 \over b} ~\log_{10} (N/ N_{\rm aft})
\label{aftervere}
\ee
This expression (\ref{aftervere}) shows that if the mainshock is taken to be the
largest event in the catalogue, then the magnitude difference  
$\langle \Delta m \rangle$ is likely to
be substantially larger than that predicted by {\it Vere-Jones}' initial
formulation. That is because the subset formed by the aftershocks is
significantly smaller than the original catalogue of size $N$, 
thus the rank of the largest aftershock is not $2$ in general.
In this sense, the mainshock in a sequence is not a member of the
set of aftershocks [{\it Utsu}, 1969; {\it Evison and Rhoades}, 2001].
The mainshock appears to be an outlier when we compare the mainshock
magnitude with the aftershock magnitude distribution, even if 
all events in the initial catalog have the same magnitude distribution.
The fact that the mainshock does not belong to the subset of
aftershocks does not however imply that mainshocks are physically different
from other earthquakes, in contradiction with previous claims of
{\it Utsu} [1969], but simply results from the rules of aftershock selection.
Expression (\ref{aftervere}) retrieves B{\aa}th's law only for a specific value 
of the number of aftershocks $N_{\rm aft} =10^{-b~\langle \Delta m 
\rangle}~10^{b(m_M-m_0)}$ with  $\langle \Delta m \rangle = 1.2$.

 We use in section \ref{ETAS} the ETAS model of seismicity,
in order to take into account the rules of aftershock selection in
time, space and magnitude, and to take into account the increase
of aftershock productivity with the maisnhock magnitude.
We will generalize the result (\ref{aftervere}) in the case where the
mainshock is not the largest event of the whole catalog.

Using an approach similar to expression (\ref{aftervere}) and taking
B{\aa}th's law as given led 
{\it Michael and Jones} [1998] and {\it Felzer et al.} [2002]  to deduce that
the number of earthquakes triggered by an earthquake of magnitude $m$ is
proportional to $\sim 10^{\alpha m}$, with $\alpha =b$.
We shall see below using numerical simulations that the ETAS model is 
also consistent with  $\alpha < b$.

\section{B{\aa}th's law and the ETAS model}
\label{ETAS}
In order to shed light on the explanation of B{\aa}th's law,
and to investigate the effects of the selection procedure for 
aftershocks, we need a complete model of seismicity, which 
describes the distribution of earthquakes in time, space 
and magnitude, and which incorporates realistic aftershock
properties. We thus study the Epidemic Type
Aftershock Sequence model (ETAS) of seismicity, introduced by
[{\it Kagan and Knopoff}, 1981; {\it Ogata}, 1988].
The ETAS model assumes that each earthquake triggers aftershocks with
a rate (productivity law) increasing as $\rho(m) = K~  10^{\alpha(m-m_0)}$
with its magnitude $m$. A crucial assumption of the ETAS model is that all 
earthquakes have the same magnitude
distribution, given by the GR law, independently of the past seismicity.
The seismicity results from the sum of an external constant average 
loading rate and from earthquakes triggered by these sources in direct
lineage or through a cascade of generations. 

It can be shown that the average total number of aftershocks 
$\langle N_{\rm aft}(m_M) \rangle$ (including the cascade of
indirect aftershocks) has the same dependence with the mainshock magnitude
\be
\langle N_{\rm aft}(m_M) \rangle =  {K \over 1-n}~10^{\alpha (m_M-m_0)}
\label{N}
\ee 
as the number of direct aftershocks $\rho(m_M)$
given above [{\it Helmstetter and Sornette}, 2002].
$n$ is the branching ratio, defined as the average number of
directly triggered earthquakes per earthquake, averaged over all magnitudes.
Using this model, {\it Felzer et al.} [2002] have argued that $\alpha$ 
must be equal to $b$ in order to obtain an average difference in magnitude 
$\langle \Delta m \rangle$  that is independent of the mainshock magnitude.
This result is in apparent disagreement with the empirical observation
$\alpha \approx 0.8 < b \approx 1$ reported by {\it Helmstetter} [2003]
using a catalog of seismicity for Southern California. 
The analysis of  {\it Felzer et al.} [2002] neglects the fluctuations
of the number of aftershocks from one sequence to another one.
We have however shown recently [{\it Saichev et al.}, 2003] that there
are huge fluctuations of the number of aftershocks per sequence for
the same mainshock magnitude. We show below, using numerical
simulations of the ETAS model, that taking into account
these fluctuations has important effects on the estimation of 
$\langle \Delta m \rangle$ and on its dependence with $m_M$.

The average magnitude $\langle m_A \rangle$ of the largest event in a 
catalog of $N_{\rm aft}$ aftershocks with magnitudes larger than  $m_0$ 
distributed according to the GR law is given by [{\it Feller}, 1966]
\ba
\langle m_A \rangle 
&=& m_0 -\int_0^1 ~{N_{\rm aft} ~(1-x)^{N_{\rm aft}-1} ln(x) \over 
b~\ln(10)}~dx
\label{mN1} \\ 
&\approx& m_0 + \log_{10}(N_{\rm aft})/b  \mbox{  \hspace{0.5cm}  
for $N_{\rm aft}\gg1$}~. 
\label{mN2}
\ea
We derive below an approximate expression for  $\langle \Delta m \rangle$
in the ETAS model, which neglects the fluctuations of $N_{\rm aft}$, i.e. which
replaces $N_{\rm aft}$ by its average value  (\ref{N}) in (\ref{mN2}).
Using this approximation, we obtain
\be
\langle \Delta m \rangle \approx { b-\alpha \over b } (m_M-m_0) -{1 \over
b} \log_{10} \Bigl({K \over 1-n} \Bigr)~.
\label{oiusfd}
\ee
This approximate relation thus predicts an increase  of $\langle 
\Delta m \rangle$ with the mainshock magnitude if $\alpha < b$.
Expression (\ref{oiusfd}) thus predicts that B{\aa}th's law is only recovered 
for $\alpha =b$ [{\it Felzer et al.}, 2002].
Using numerical simulations of the ETAS model, we find however that
the large fluctuations in aftershock numbers due to the cascades of
triggered events modify significantly the prediction (\ref{oiusfd}).
Adding the constraint that aftershocks are usually chosen
to be smaller than the mainshock further alters the prediction (\ref{oiusfd}).

We have generated synthetic catalogs with the ETAS model to 
measure  $\langle \Delta m \rangle$
for different values of the mainshock magnitude. In this first test,
we starts the simulation with a mainshock of magnitude $m_M$,
which generates a cascade of direct and indirect aftershocks.
We select as ``aftershocks'' all earthquakes triggered directly or
indirectly by the mainshock, without any constraint in the time,
location, or magnitude of these events.
For $\alpha=0.8$ and $b=1$, we find that $\langle \Delta m \rangle$ is
much larger than predicted by  (\ref{oiusfd}), and increases slower
with the mainshock magnitude, in better agreement with B{\aa}th's law
than the analytical solution  (\ref{oiusfd}) (Figure {\ref{fig1}).
We have checked that the average number of aftershocks 
$\langle N_{\rm aft}(m_M) \rangle$ is in
good agreement with the analytical solution (\ref{N}), and thus 
a discrepancy with (\ref{N})
is not an explanation for the difference between the results of the 
numerical simulations and the analytical prediction   (\ref{oiusfd}).

The large fluctuations of the total number of aftershocks are at the 
origin of the discrepancy between the observed $\langle \Delta m
\rangle$ and the prediction
(\ref{oiusfd}), which neglects the fluctuations of the number of aftershocks.
{\it Saichev et al.} [2003] have recently demonstrated that the total
number of aftershocks in the ETAS model in the regime $\alpha>b/2$ 
has an asymptotic power-law distribution in the tail with an exponent 
of the cumulative distribution smaller than $1$, even in the
subcritical regime (defined by a branching ratio  $n<1$). These
huge fluctuations arise from the cascades of triggering and from the
power-law distribution of the number of triggered earthquakes per
triggering earthquake appearing as a combination of the GR law and of 
the productivity law $\rho(m)$.  Practically, this means  that the 
aftershock number fluctuates widely from realization 
to realization and the average will be controlled by a few
sequences that happen to have an unusually large number of aftershocks.
Numerical simulations show that $\langle \Delta m \rangle$ is not controlled
by the average number of aftershocks, but by its ``typical'' value,
which is much smaller than the average value. Therefore, the
expression (\ref{oiusfd}) of  $\langle \Delta m \rangle$ obtained by
replacing $N_{\rm aft}$ by its average value  (\ref{N}) in (\ref{mN1}) 
is a very bad approximation.
Using the exact distribution of the number of aftershocks given in
[{\it Saichev et al.}, 2003], we can obtain the asymptotic expression for
large $m_M$, which recovers the dependence of  $\langle \Delta m \rangle$
with $m_M$ predicted by  (\ref{oiusfd}).

For large mainshock magnitudes, the relative fluctuations of the total
number of aftershocks per mainshock are weaker. Therefore, the obtained
average magnitude difference $\langle \Delta m \rangle$ tends
to recover the linear dependence (\ref{oiusfd}) with the mainshock magnitude,
represented by the continuous line in Figure \ref{fig1}. Our numerical simulations
show that a constant value of  $\langle \Delta m \rangle$ in a wide range of magnitudes
can be reproduced using the ETAS model if $\alpha < b$.
Our results also predict that B{\aa}th's law should
fail for large mainshock magnitudes according to (\ref{oiusfd}) if $\alpha$ is
smaller than $b$. It is however doubtful that this deviation
from B{\aa}th's law can be observed in real data as the number of large mainshocks
is small.

\begin{figure}
\begin{center}
\includegraphics[width=8cm]{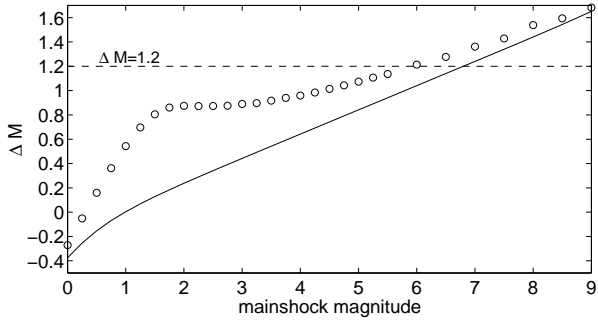}
\caption{\label{fig1} Average magnitude difference $\langle \Delta m \rangle$ 
between the mainshock and its largest aftershock as a function of 
the mainshock magnitude (open circle), for numerical simulations of 
the ETAS model with parameters $n=0.8$, $\alpha=0.8$, $b=1$, $m_0=0$ 
and with an Omori exponent $p=1.2$. The continuous line is the
prediction using the approximate 
analytical solution (\ref{oiusfd}) for  $\langle \Delta m  \rangle$.}
\end{center}
\end{figure}

While the impact of fluctuations in the number of aftershocks
produces a value of  $\langle \Delta m \rangle $ larger than
predicted by (\ref{oiusfd}) and in better agreement with  B{\aa}th's law, 
the average magnitude difference $\langle \Delta m \rangle \approx 0.7$ 
remains smaller than the empirical value  $\langle \Delta m \rangle \approx 1.2$. 
However, we have not yet taken into
account the constraints of aftershocks selection, which will
further modify  $\langle \Delta m \rangle $. In the simulations
giving Figure \ref{fig1}, all earthquakes triggered (directly or
indirectly) by the mainshock have been considered as aftershocks
even if they were larger than the mainshock.
In real data, the difficulty of identifying aftershocks and the
usual constraint that aftershocks are smaller than the mainshock can be
expected to affect the relation between $\langle \Delta m \rangle$
and the mainshock magnitude.
The selection of aftershocks requires the choice of a space-time
window to distinguish aftershocks from background events.
A significant fraction of aftershocks can thus be missed. As
a consequence, the value of $\Delta m$ will increase.

In order to quantify the impact of these constraints, we have generated
synthetic catalogs using the ETAS model, which include a realistic
spatio-temporal distribution of aftershocks. 
Specifically, according to the ETAS model,
the number of aftershocks triggered directly by an event
of magnitude $m$, at a time $t$ after the mainshock and at a distance
$r$ is given by
\be
\phi_m(t,r) = n~{(b-\alpha)\over b}~10^{\alpha(m-m_0)}~
{\theta~~ c^{\theta} \over (t+c)^{p}}
~{\mu ~~ {d_m}^{\mu} \over (r+d_m)^{1+\mu}}~.
\label{phi}
\ee
where $n$ is the branching ratio, $p$ is the exponent 
of the local Omori's law  (which is generally larger than the observed
 Omori exponent)  and $d_m$ is the characteristic size 
of the aftershock cluster of a magnitude $m$ earthquake given by 
$d_m=0.01 \times 10^{0.5  m}$~km.

We have then applied standard rules for the selection of aftershocks. 
We consider as a potential mainshock 
each earthquake that has not been preceded by a larger earthquake in a
space-time window  $R_C \times T_c$. This rule allows us to remove the 
influence of previous earthquakes and to obtain an estimate of the rate of
seismicity triggered by this mainshock. The constant $R_c$ is fixed equal to 
the size $\approx 100$~km of the largest cluster in the catalog and 
$T_c=100$~days. We then define aftershocks as all events occurring in a 
space time window $R(m_M) \times T(m_M)$ after a mainshock of magnitude $m_M$, 
where both $R(m_M)=2.5 \times 10^{(1.2m_M-4)/3}$~km and 
$T(m_M)= 10/3 \times 10^{(2/3)(m_M-5)}$~days increase with the mainshock 
magnitude $m_M$ [{\it Kagan}, 1996; {\it Console et al.}, 2003].

\begin{figure}
\begin{center}
\includegraphics[width=8cm]{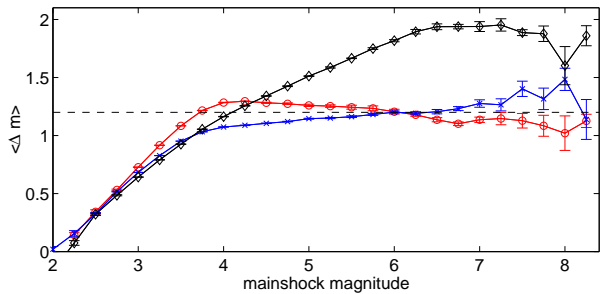}
\caption{\label{fig2} Average magnitude difference $\langle \Delta m \rangle$
between a mainshock and its largest aftershock, for numerical
simulations of the ETAS model with parameters $b=1$, $c=0.001$ day,
$p=1.2$, a minimum magnitude $m_0=2$, a maximum magnitude 
$m_{max}=8.5$ and a constant loading $\mu=300$ events per day. 
Each curve corresponds to a different value
of the ETAS parameters: $\alpha=0.8$ and $n=0.76$ (crosses), 
$\alpha=0.5$ and $n=0.8$ (diamonds) and $\alpha=1$ and $n=0.6$ (circles). 
The error bars gives the uncertainty of  $\langle \Delta m \rangle$
(1 standard deviation). The horizontal dashed line is the empirical
value $\langle \Delta m \rangle=1.2$.}
\end{center}
\end{figure}

The results for different values of $\alpha$ are represented in 
Figure~\ref{fig2}. For intermediate mainshock magnitude, the average magnitude 
difference $\langle \Delta m \rangle$ for $\alpha=0.8$ is significantly 
larger than found in Figure~\ref{fig1} without the selection procedure,
because mainshocks which trigger a larger event are rejected, and because
the rules of selection (with a time-space window $R(m)$ and  $T(m)$
increasing with $m$) reject a large number of aftershocks, especially for small
mainshocks. For small magnitude $m_M$,   $\langle \Delta m 
\rangle$ is small and then increases rapidly with $m$.
This regime is not pertinent because most mainshocks do not 
trigger any aftershock and are thus rejected from the analysis. 
Most studies have considered only mainshocks  with magnitude $m \geq m_0+2$,
where $m_0$ is the minimum detection threshold.
For $\alpha=0.8$ or $\alpha=1$, the magnitude difference is 
$\approx 1.2$ in a large range of mainshock magnitudes, 
in agreement with B{\aa}th's law. 
For $\alpha=1$, there is a slight decrease of $\langle \Delta m 
\rangle$ with $m_M$.
For $\alpha=0.5$, we observe a fast increase of $\langle \Delta m 
\rangle$ with $m_M$, which is not consistent with the observations
of B{\aa}th's law  [e.g. {\it Felzer et al.}, 2002; {\it Console et al.}, 2003].
The shape of the curves $\langle \Delta m \rangle$ is mostly 
controlled by $\alpha$.
The other parameters of the ETAS model and the rules of aftershock
selection increase or decrease  $\langle \Delta m \rangle$ but do
not change the scaling of  $\langle \Delta m \rangle$  with the 
mainshock magnitude.

\section{Discussion and conclusion}

We have first shown that the standard interpretation of B{\aa}th's law
in terms of the two largest events of a self-similar set of
independent events is incorrect.
We have stressed the importance of the selection process of aftershocks,
which represent only a subset of the whole seismicity catalog.
Our point is that the average magnitude difference 
$\langle \Delta m \rangle$ is not only controlled by the magnitude 
distribution but also by the aftershock productivity.
A large magnitude difference  $\langle \Delta m \rangle$ can be explained
by a low aftershock productivity.

Using numerical simulations of the ETAS model, we have shown that
this model is in good agreement with B{\aa}th's law in a certain range of
the model parameters.
We have pointed out the importance of the selection process of aftershocks,
of the constraint that aftershocks are smaller than the mainshock and of
the fluctuation of the number of aftershocks per sequence
in the determination of the value of $\langle \Delta m \rangle$,
and in its apparent independence as a function of the mainshock magnitude.
In the ETAS model, the cascades of multiple triggering
induce large fluctuations of the total number of aftershocks.
These fluctuations in turn induce a modification of the scaling
of $\langle \Delta m \rangle$ with the mainshock magnitude by 
comparison with the predictions neglecting these fluctuations. 
The constraints due to aftershock selection further affect the value
of $\langle \Delta m \rangle$. 
Observations that $\langle \Delta m \rangle$ does not vary
significantly with the mainshock magnitude requires that the exponent of 
the aftershock productivity law is in the range $0.8<\alpha<1$.
B{\aa}th's law is thus consistent with the regime $\alpha<b$ in which
earthquake triggering is dominated by the smallest earthquakes [{\it
Helmstetter}, 2003].

\begin{acknowledgments}
This work is partially supported by NSF-EAR02-30429, by
the Southern California Earthquake Center (SCEC) and by
the James S. Mc Donnell Foundation 21st century scientist
award/studying complex system. 
SCEC is funded by NSF Cooperative Agreement EAR-0106924 and USGS Cooperative
Agreement 02HQAG0008.  The SCEC contribution number for this paper is 739. 
We acknowledge useful discussions with R. Console, D. Jackson, 
Y. Kagan and D. Vere-Jones.
\end{acknowledgments}

}
\end{article}
\end{document}